\documentclass[namedreferences]{SolarPhysics}
\usepackage[optionalrh]{spr-sola-addons} 
\usepackage{epsfig}
\usepackage{graphicx}
\usepackage{color}
\usepackage{url}
 
\begin{document}
\begin{article}
\begin{opening}

\title{Predicting the 
 Amplitude of a Solar Cycle  Using the North-South Asymmetry 
in the Previous Cycle: II. An Improved Prediction for Solar Cycle~24}

\author{J. JAVARAIAH$^{1,2}$}

\runningauthor{J. JAVARAIAH}
\runningtitle{PREDICTION OF THE AMPLITUDE OF A SOLAR CYCLE}

\institute{(1)\  Indian Institute of Astrophysics, Bangalore-560 034, India.\\
email: \url{jj@iiap.res.in}\\          
(2)\ Department of Physics and Astronomy,
  UCLA, 430 Portola Plaza,  Los Angeles, CA 90095-1547, U. S. A.\\ 
}

\begin{abstract}
Recently,
using   Greenwich and Solar Optical Observing Network sunspot group data during 
the period 1874\,--\,2006,  Javaraiah ({\it Mon. Not. Roy. Astron. Soc.}  377, L34, 2007: Paper~I),    has  found that:
 (1)  the sum of the areas of the sunspot groups  in
 $0^\circ$\,--\,$10^\circ$ latitude  
 interval of  the Sun's  northern hemisphere and in the time-interval  
of $-1.35$ year to $+2.15$ year from the time of the preceding minimum
 of a solar cycle $n$ correlates well (corr. coeff. $r = 0.947$) 
 with the  amplitude
(maximum of the smoothed monthly sunspot number) of 
the next cycle $n+1$. 
 (2)  The sum of the areas of the spot groups  in   
$0^\circ$\,--\,$10^\circ$ latitude 
 interval of the 
 southern hemisphere and in the time-interval of 
 1.0 year to 1.75 year just after the time of the maximum of the cycle $n$  
 correlates very well
 ($r = 0.966$) 
with the amplitude of cycle $n+1$.    
Using these relations, (1) and (2),   
the values  
$112 \pm 13$ and $74 \pm 10$, respectively, were  predicted in Paper~I 
for the amplitude of the 
upcoming cycle 24.  
Here we found that  
 the  north-south asymmetries in the aforementioned   area sums have a strong 
$\approx$ 44-year periodicity and  from   
 this  we can infer 
that the upcoming cycle 24 will be weaker than cycle~23. 
In case of  (1),   
 the  
north-south asymmetry in the  area sum of a cycle $n$ 
  also has a  
 relationship, say (3), with the amplitude of cycle~$n+1$, which is similar
  to (1) but  more statistically significant 
($r=0.968$) like
 (2).
 By using  (3)  it is  possible to  predict 
the amplitude of a cycle with a better accuracy  
 by about 13 years in advance, and we get   $103 \pm 10$ for    
the amplitude of the upcoming 
cycle~24.  However, we found a similar 
but a more
 statistically significant 
    ($r = 0.983$) relationship, say (4), by using  
 the sum of the area sum used in (2) and the north-south 
difference  used in  (3).   
 By using (4) it is possible to predict
 the amplitude of a cycle 
by about 9 years in advance with a high accuracy and    
  we get  $87 \pm 7$  for the amplitude of cycle~24, 
which is about 28\% less than the amplitude of cycle~23.
Our results  also indicate that  cycle~25 will be
 stronger than cycle~24.
 The  variations in the mean  
meridional motions of the spot groups during  odd and even numbered cycles
 suggest that 
 the solar meridional 
flows may transport  
 magnetic flux across the solar equator and 
potentially responsible for  all the above relationships.   
\end{abstract}

\end{opening}
%________________________________________________________________

\section{Introduction}

Prediction of the strength of a solar cycle well in advance is important 
for predicting the space weather because 
solar activity affects space weather in several ways~\cite{hw04,hw06,kane07a}. 
In addition,  
it may help for understanding the basic physical processes responsible 
for solar cycle~\cite{ddg06,ccj07,cs07,jcc07}.  
Many  characteristics of a solar cycle  
and of its previous cycles  
have been used to predict the amplitudes of
 the cycle~\cite{hwr99,li01,kane07a,obrid08}. 
The existence of the north-south asymmetry in solar activity 
is well known since several  decades and  has been extensively studied  
using the data on  
almost all the solar activity phenomena including sunspot number, 
sunspot groups, solar flares, prominences/filaments, photospheric 
magnetic fields, solar rotation and differential rotation, 
meridional flow, etc., and revealed 
many characteristics of it~\cite{roy77,sks86,ahs90,garcia90,hw90,yi92,cob93,verma93,sokn94,ao96,jg97,dd96,li02,jj03,georg03,georg05,boc05,gjk05,ksb05,jp05,
temm06,za06,ju06,chang08}.
However, so far no  physical or even a significant 
statistical  relationship between 
the solar activity and its north-south asymmetry is known. 
Recently, Javaraiah (2007, hereafter Paper~I) 
 has 
 found   
 the following  statistically high significant 
relationships 
 between 
 the sums of the areas 
 of the sunspot groups, $A_{\rm N}$ and $A_{\rm 
 S}$ (normalized by 1000 ), in    
$0^\circ$\,--\,$10^\circ$ 
latitude interval of the  
 Sun's northern hemisphere and in  
  the time interval of 
  $T_{\rm m}^*: T_{\rm m} +  (- 1.35\ {\rm to}\ + 2.15)$, $i.e.$ 
 $-1.35$ year to $+2.15$ year  
 from the time of the preceding minimum ($T_{\rm m}$) - and  
in the same latitude interval of the  
 southern hemisphere but in  
  the time interval of 
  $T_{\rm M}^*: T_{\rm M} + (1.0\ {\rm to}\  1.75)$, $i.e.$ 1.0 year to 1.75 year    
just after the time of the maximum ($T_{\rm M}$)- during   a solar
 cycle $n$ and  
 the amplitude ($R_{\rm M}$, maximum of the smoothed monthly sunspot number) of 
the next cycle $n+1$:
$$R_{{\rm M}, n+1} = (1.72 \pm 0.19) A_{{\rm N}, n} (T_{\rm m}^*) + (74.0 \pm 7.0) , \eqno(1)$$
$$R_{{\rm M}, n+1} = (1.55 \pm 0.14) A_{{\rm S}, n} (T_{\rm M}^*) + (21.8 \pm 9.6) , \eqno(2)$$

\noindent where   $n = 12,\ 13, ....., 23$ represents the Waldmeier solar cycle 
number.
The corresponding  correlation coefficients of these relations are
  $r =   0.947$ 
 and   $r =   0.966$, respectively.
By using Equations~(1) and (2)  
 in Paper~I we have found $112 \pm 13$ and $74 \pm 10$, respectively, 
for  $R_{\rm M}$ of the upcoming solar 
cycle~24.  Since Equation~(2) is  statistically 
more significant than  Equation~(1), hence,   
in Paper~I we have  predicted the value $74 \pm 10$  for $R_{\rm M}$ of 
the upcoming cycle~24. 
In the present paper we have made a detailed study on the  
 north-south asymmetries in the sums of the 
areas of the spot groups, $i.e.$,  differences   between  
the $A_{{\rm N}, n}$  and $A_{{\rm
 S}, n}$,  during both the time intervals  $T_{\rm m}^*$ and $T_{\rm M}^*$ 
of  the 
solar cycles $n = 12$ to $23$. This enabled us to find    
two more new  relationships, (3) and (4), and using these to  improve 
the  prediction  in Paper~I for the amplitude of  solar cycle~24. 
The  physical significance of all the aforementioned 
relationships is discussed.

It should be noted here that in most of the studies
of the north-south asymmetry of 
a solar activity phenomenon, the north-south asymmetry 
 is determined using the conventional 
formula, $\frac{N-S}{N+S}$, where 
$N$ and $S$ are the corresponding 
quantities of the  activity phenomenon in the northern and the 
southern hemispheres, respectively. This fraction has a strong 
11\,--\,12 year periodicity~\cite{cob93}.  \inlinecite{yi92}  
pointed out that in the north-south asymmetry, derived by using this formula,   
the 11\,--\,12 year periodicity  
has no statistical significance because  mostly it is an artifact of
 the 11-year 
cycle of $N+S$. In order to verify this \inlinecite{jg97} 
 determined the power spectra of both the  $N-S$ and
  the $\frac{N-S}{N+S}$ of the  sunspot number data and found that 
a peak at 11\,--\,12 year in the 
power spectrum of $N-S$ is statistically  insignificant, whereas it is 
statistically very significant in the spectrum of $\frac{N-S}{N+S}$, confirming 
the doubt of~\inlinecite{yi92}. Recently, \inlinecite{boc05} also 
confirmed the same. Therefore,  the difference $N-S$  seems to 
 represent the   
north-south asymmetry of a solar activity 
phenomena more appropriately  than the 
 fraction $\frac{N-S}{N+S}$. 
In addition, in the present analysis we found that the 
correlation ($r =0.90$) between $\frac{A_{\rm N}-A_{
\rm S}}{A_{\rm N}+A_{
\rm S}}$ of a cycle $n$ 
and the $R_{\rm M}$ of cycle 
$n+1$ is much weaker than that ($r =0.968$) 
between the difference $A_{\rm N}-A_{
\rm S}$ during $T_{\rm m}^*$ of a cycle $n$ and the $R_{\rm M}$ of 
the cycle $n+1$. Therefore, here we have used the difference $A_{\rm N}-A_{
\rm S}$. 

Since a large number of abbreviations are used here, 
hence for the sake of the readers 
convenience 
 we listed below the  
 meanings of all the abbreviations
(thanks are due to  the anonymous referee's suggestion):
\begin{itemize}
\item $n$ - the Waldmeier solar cycle number,
\item $T_{\rm m}$ - the  preceding  minimum epoch of a solar cycle,
\item $T_{\rm M}$ - the maximum epoch of a solar cycle,
\item $R_{\rm m}$ - the value of smoothed monthly sunspot number in $T_{\rm m}$,
\item $R_{\rm M}$ - the value of smoothed monthly number  in $T_{\rm M}$,
\item $T_{\rm m}^*$ - the time-interval of $-1.35$ year to $+2.15$ year from $T_{\rm m}$,
\item $T_{\rm M}^*$ - the time-interval of $1.0$ year to $1.75$ year just after $T_{\rm M}$,
\item $A_{{\rm N}, n} (T_{\rm m}^*)$ - the sum of the areas of spot groups in  
 $0^\circ - 10^\circ$ latitude interval of the northern hemisphere 
 during $T_{\rm m}^*$ of a cycle $n$,  
\item $A_{{\rm
 S}, n} (T_{\rm m}^*)$ - the sum of the areas of spot groups in  
 $0^\circ - 10^\circ$ latitude interval of the southern  hemisphere 
during $T_{\rm m}^*$ of a cycle $n$,  
\item $A_{{\rm N}, n} (T_{\rm M}^*)$ - the sum of the areas of spot groups in  
$0^\circ - 10^\circ$ latitude interval of the northern hemisphere 
during $T_{\rm M}^*$ of a cycle $n$,  
\item $A_{{\rm 
 S}, n} (T_{\rm M}^*)$ - the sum of the areas of spot groups in  
$0^\circ - 10^\circ$ latitude interval of the southern hemisphere 
during $T_{\rm M}^*$ of a cycle $n$, 
\item $\delta A_n$ -  the difference $A_{{\rm N}, n} (T_{\rm m}^*) - A_{{\rm
 S}, n} (T_{\rm m}^*)$,
\item $A_{{\rm NS}, n}$ -  the sum  $A_{{\rm N}, n} (T_{\rm m}^*) + A_{{\rm 
 S}, n} (T_{\rm M}^*)$, and
\item $\Delta_{\delta{A{{\rm   
 S}, n}}}$ - the sum  $\delta A_n (T_{\rm m}^*) + A_{{\rm 
 S}, n} (T_{\rm M}^*)$.
\end{itemize} 

In the next section we  describe the data analysis and the results. In 
Section~3 we 
  discuss  about the implications of 
 the relationships among $A_{{\rm N}, n} (T_{\rm m}^*)$, $A_{{\rm 
 S}, n} (T_{\rm M}^*)$, ${\delta A}_n (T_{\rm m}^*)$,   
 $\Delta_{\delta{A{{\rm S}, n}}}$, and $R_{{\rm M}, n+1}$ for understanding the underlying 
physical process of the solar cycle. For this purpose we have determined the 
 variations in the mean meridional motions 
of the spot groups 
in the northern and the southern hemispheres
 during the odd numbered cycles (11, 13, 15, 17, 19, 21, 23) and the even numbered 
cycles (12, 14, 16, 18, 20, 22),  using 
 the superposed epoch analysis as described in \inlinecite{ju06}. 
In the same section 
 we  compare our prediction with  
other  authors' 
 predictions for the $R_{\rm M}$ of solar cycle~24, particularly   
 with those predictions 
  based on the flux-transport-dynamo models,  and the spectral analyses 
and 
magnetic oscillations models.
In Section~4, we have given the summary of the results and the conclusions.

\section{Data Analysis and Results}

\subsection{Data}

Here we have used the values of $A_{\rm N}$ and $A_{
\rm S}$, $i.e.$,  
the sums of the areas of the sunspot groups  
in $0^\circ-10^\circ$ latitude intervals of the northern   
and the southern 
hemispheres  during  the time intervals 
$T_{\rm m}^*$ and 
$T_{\rm M}^*$ of cycles  12\,--\,23, which were  determined in Paper I by 
using 
the Greenwich sunspot group data 
during the period 1874\,--\,1976, and 
the sunspot group data from the Solar Optical 
Observing Network (SOON)  of the US Air Force/US National Oceanic and 
Atmospheric Administration  during 1977 January 1\,--\,2005 September 30.  
The recently updated these data were taken   from 
the NASA web-site of David Hathaway
 ({\tt http://solarscience.msfc.nasa.gov/greenwch.shtml}). 
 In Table~1 we have given the values of $A_{\rm N}$ and $A_{
\rm S}$ during both  
the time intervals 
 $T_{\rm m}^*$ and $T_{\rm M}^*$, and  
   the maximum ($R_{\rm M}$) and the minimum 
($R_{\rm m}$) amplitudes 
(the largest and   the smallest  smoothed monthly mean sunspot numbers)
and the corresponding epochs $T_{\rm M}$ and $T_{\rm m}$   
of the  solar-cycles 12\,--\,23 
(the values of $T_{\rm m}$, $T_{\rm M}$, $R_{\rm m}$ and $R_{\rm M}$ of cycles~12\,--\,23 are
 taken  from the website, 
{\tt ftp://ftp.ngdc.noaa.gov/STP\break /SOLAR\_DATA/SUNSPOT\_NUMBERS}). 
 The 
details of the data reduction and the analysis were described in Paper~I.
The method of determination of $A_{\rm N} (T_{\rm m}^*)$ and $A_{
\rm S} (T_{\rm M}^*)$,  as described 
in Paper I, is as follows:  
 first we determine the sums of the daily areas of the spot groups
 in the time-intervals whose lengths are chosen arbitrarily 
 around the minimum and the maximum of a cycle $n$  
 and $R_{\rm M}$ of the  cycle  $n+1$ or $n+2$, etc.,  and then we 
determine the exact time intervals  
 by increasing or  decreasing the arbitrary intervals  
with 0.05 year until we get the   
maximum correlation. 
We have chosen this method 
because the lengths of the solar  cycles vary.  However,   
 it may   be possible to obtain the similar results 
by  binning the available  data for the entire 
period (1974\,--\,2007) into successive intervals of suitable sizes 
 and then picking the  intervals 
close to  the minimum and maximum epochs of the cycles. 

Note: In case of  $0^\circ -10^\circ$ latitude interval  of 
the southern hemisphere,   a reasonably good 
correlation is found between the sum of the areas of the spot groups during  
a long  time-interval approximately one year after the maximum to approximately 
two year 
before the end of cycle $n$ and   $R_{\rm M}$ of cycle $n+1$. Since  the correlation 
is maximum  for $T_{\rm M}^*$, hence we have considered it.  Although  
 $T_{\rm M}^*$ is much shorter than $T_{\rm m}^*$,  
$A_{
\rm S} (T_{\rm M}^*)$ is considerably larger than $A_{\rm N} (T_{\rm m}^*)$.  Hence, in spite of 
 $T_{\rm M}^*$ is short the 
 $A_{
\rm S} (T_{\rm M}^*)$ time series is also well defined.

\begin{table*} 
\flushleft
{\scriptsize
   \caption{The maximum ($R_{\rm M}$) and the minimum 
($R_{\rm m}$) amplitudes 
(the largest and the smallest  smoothed monthly mean sunspot numbers)
of the  solar-cycles $n =$ 12\,--\,23 and   
the sums of the areas of spot groups  (normalized by 1000)
in $0^\circ-10^\circ$ latitude intervals of the northern hemisphere ($A_{\rm N}$) 
and the southern 
hemisphere ($A_{
\rm S}$) during  the time intervals 
$T_{\rm m}^* = T_{\rm m} + (-1.35\ {\rm to}\ 2.15)$ and 
$T_{\rm M}^* = T_{\rm M} + (1.0\ {\rm to}\ 1.75)$, 
where $T_{\rm M}$ and $T_{\rm m}$ represent  the maximum
 and the preceding minimum epochs of the 
solar cycles, respectively.}

\begin{tabular}{lccccccccccccccccc}
\hline
  \noalign{\smallskip}
Cycle & \multicolumn{2}{c}{Minimum} & \multicolumn{2}{c}{Maximum} & 
\multicolumn{3}{c}{Around minimum}& \multicolumn{3}{c}{After maximum} \\
  \noalign{\smallskip}

$n$&$T_{\rm m}$&$R_{\rm m}$&$T_{\rm M}$&$R_{\rm M}$&$T_{\rm m}^*$&$A_{\rm N}$&$A_{\rm  S}$&$T_{\rm M}^*$&$A_{\rm N}$&$A_{\rm S}$\\
\hline
  \noalign{\smallskip}

12&1878.9&2.2 &1883.9& 74.6 &1877.55\,--\,1881.05& 9.47&9.84&1884.90\,--\,1885.65&41.38&42.11\\
13&1889.6&5.0 &1894.1& 87.9 &1888.25\,--\,1891.75& 3.22&30.83&1895.10\,--\,1895.85&18.67&32.64\\
14&1901.7&2.6 &1907.0& 64.2 &1900.35\,--\,1903.85&12.98&10.94&1908.00\,--\,1908.75&35.75&54.64\\
15&1913.6&1.5 &1917.6&105.4 &1912.25\,--\,1915.75& 3.74&7.93&1918.60\,--\,1919.35&84.76&34.58\\
16&1923.6&5.6 &1928.4& 78.1 &1922.25\,--\,1925.75&33.96&13.03&1929.40\,--\,1930.15&64.20&75.96\\
17&1933.8&3.4 &1937.4&119.2 &1932.45\,--\,1935.95&29.96&7.43&1938.40\,--\,1939.15&49.12&82.01\\
18&1944.2&7.7 &1947.5&151.8 &1942.85\,--\,1946.35&69.35&12.06&1948.50\,--\,1949.25&70.31&119.65\\
19&1954.3&3.4 &1957.9&201.3 &1952.95\,--\,1956.45&15.23&13.18&1958.90\,--\,1959.65&116.12&53.01\\
20&1964.9&9.6 &1968.9&110.6 &1963.55\,--\,1967.05&50.31&6.89&1969.90\,--\,1970.65&37.96&78.28\\
21&1976.5&12.2&1979.9&164.5 &1975.15\,--\,1978.65&60.05&25.06&1980.90\,--\,1981.65&58.42&83.53\\
22&1986.8&12.3&1989.6&158.5 &1985.45\,--\,1988.95&29.85&20.93&1990.60\,--\,1991.35&41.33&67.48\\
23$^{\mathrm{a}}$&1996.4&8.0 &2000.3&120.8 &1995.05\,--\,1998.55&21.99&20.17&2001.30\,--\,2002.05&77.27&33.58\\
\hline
  \noalign{\smallskip}

\end{tabular}

$^{\mathrm{a}}$ indicates the incompleteness of the current cycle~23.
}
\end{table*}

\subsection{Variations in the Sums of the Areas of the Spot Groups} 
Figure~1 shows the  variations in   
$A_{\rm N} (T_{\rm m}^*)$, $A_{
\rm S} (T_{\rm m}^*)$, $A_{\rm N} (T_{\rm M}^*)$  and $A_{
\rm S} (T_{\rm M}^*)$ 
during cycles~12\,--\,23. 
In this figure
we have also shown the variation in $R_{\rm M}$ of  cycles~12\,--\,23.   
In Figure~1(a), it can be seen that there exist considerable  differences 
between  $A_{\rm N} (T_{\rm m}^*)$ and $A_{
\rm S} (T_{\rm m}^*)$. 
The amplitude of the cycle-to-cycle variation 
of  $A_{\rm N} (T_{\rm m}^*)$   is much  larger than 
that of the corresponding variation of $A_{
\rm S} (T_{\rm m}^*)$.
There  is a strong suggestion of the
existence of a periodicity of about $\approx$44 years (double Hale cycle) 
in $A_{\rm N} (T_{\rm m}^*)$.
 $T_{\rm m}^*$ of the largest amplitude cycle 19 is one of the   
 minimum epochs of the $\approx$44 years  cyclic variation.
  $A_{\rm N} (T_{\rm m}^*)$ leads $R_{\rm M}$ by about 13 years [as already known 
from the relationship,
Equation~(1) above].
 The  cycle-to-cycle variation in $A_{
\rm S} (T_{\rm m}^*)$ 
 is weak [$A_{
\rm S} (T_{\rm m}^*)$ is almost constant from cycle 14 to cycle 20], but    
 there is an evidence for the existence of a weak  
  80\,--\,90 year cycle (Gleissberg cycle). 
In Figure~1(b) it can be seen that there 
is no much difference between the maximum amplitudes of the variations
in $A_{\rm N} (T_{\rm M}^*)$ and $A_{
\rm S} (T_{\rm M}^*)$.
As we  already know from Paper~I   
the pattern of $A_{
\rm S} (T_{\rm M}^*)$  is similar to that of $A_{\rm N} (T_{\rm m}^*)$ ($r = 0.94$), 
and  $A_{
\rm S} (T_{\rm M}^*)$ 
leads $R_{\rm M}$ by 
 about 9\,--\,10 years.  
 $A_{
\rm S} (T_{\rm M}^*)$ has a strong $\approx$ 44-year periodicity.  
   $A_{\rm N} (T_{\rm M}^*)$   varies  
approximately in phase with $R_{\rm M}$ ($r= 0.6$) and 
has a somewhat different periodicity,   33\,--\,44 year (33-year periodicity 
associates with the high level of activity). 
 The  correlation between 
$A_{\rm N} (T_{\rm M}^*)$  and $R_{\rm M}$ indicates that there is an influence
 of the latter  on the former. It should be noted here that $R_{\rm M}$ of a cycle $n$ leads 
$A_{\rm N} (T_{\rm M}^*)$ of the same cycle by 1.0 year to 1.75 year.
There is no significant correlation either between $A_{\rm N} (T_{\rm m}^*)$ and 
$A_{
\rm S} (T_{\rm m}^*)$ or between $A_{\rm N} (T_{\rm M}^*)$ and $A_{
\rm S} (T_{\rm M}^*)$.

\begin{figure}
\centerline{\includegraphics[width=12.0cm]{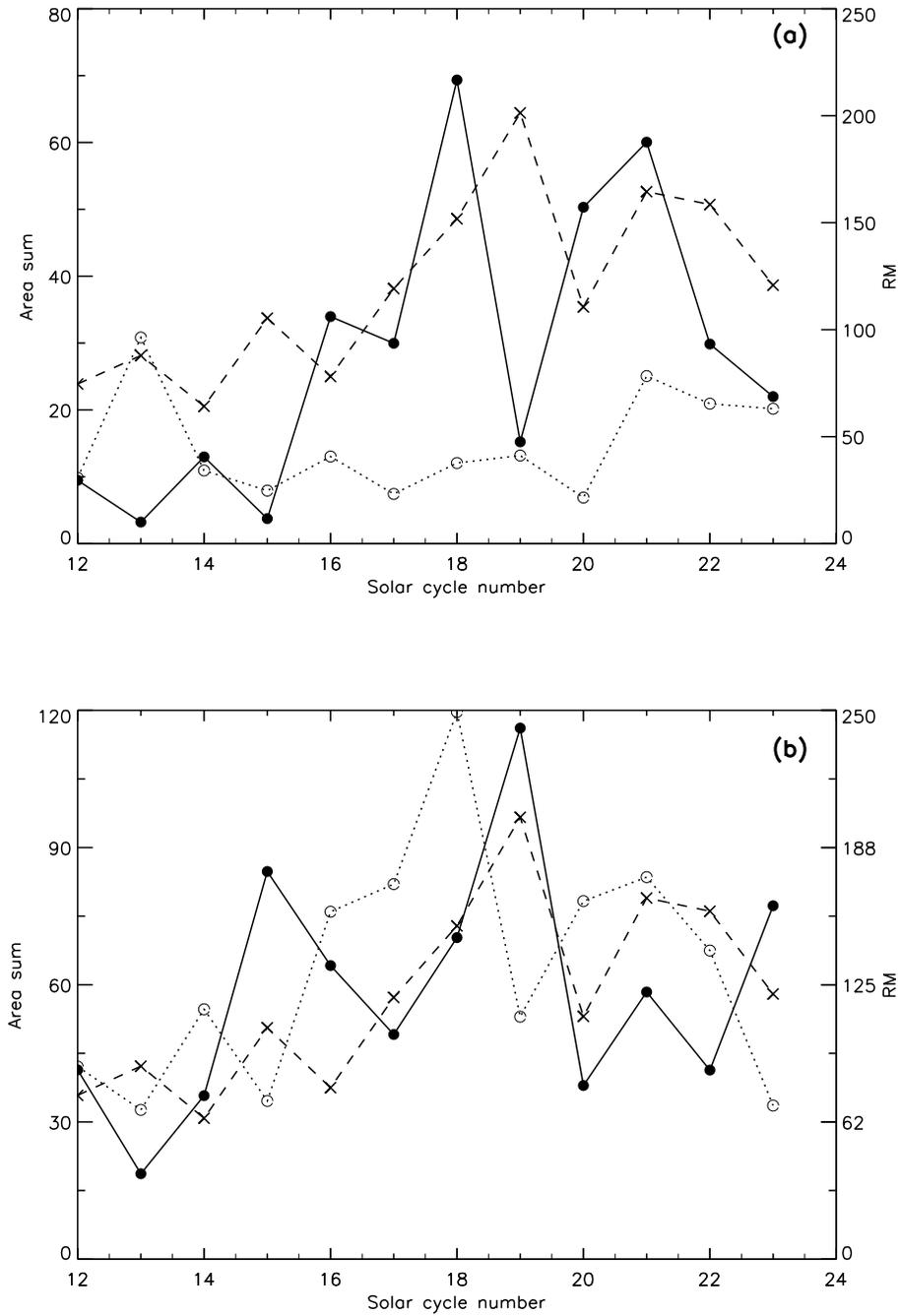}}
\caption{Plots of  the sums of the areas of 
sunspot groups - $A_{\rm N}$ (filled circle-solid curve) and $A_{\rm S}$
(open circle-dotted curve)     
in $0^\circ-10^\circ$ latitude intervals of the northern 
 and the southern hemispheres, respectively,
 during  (a) $T_{\rm m}^*$ and (b) $T_{\rm M}^*$ - versus solar cycle number.
  In both (a) and (b)
the  cross-dashed curve  represents
 the variation in $R_{\rm M}$.} 
\end{figure}

\begin{figure}
\centerline{\includegraphics[width=12.0cm]{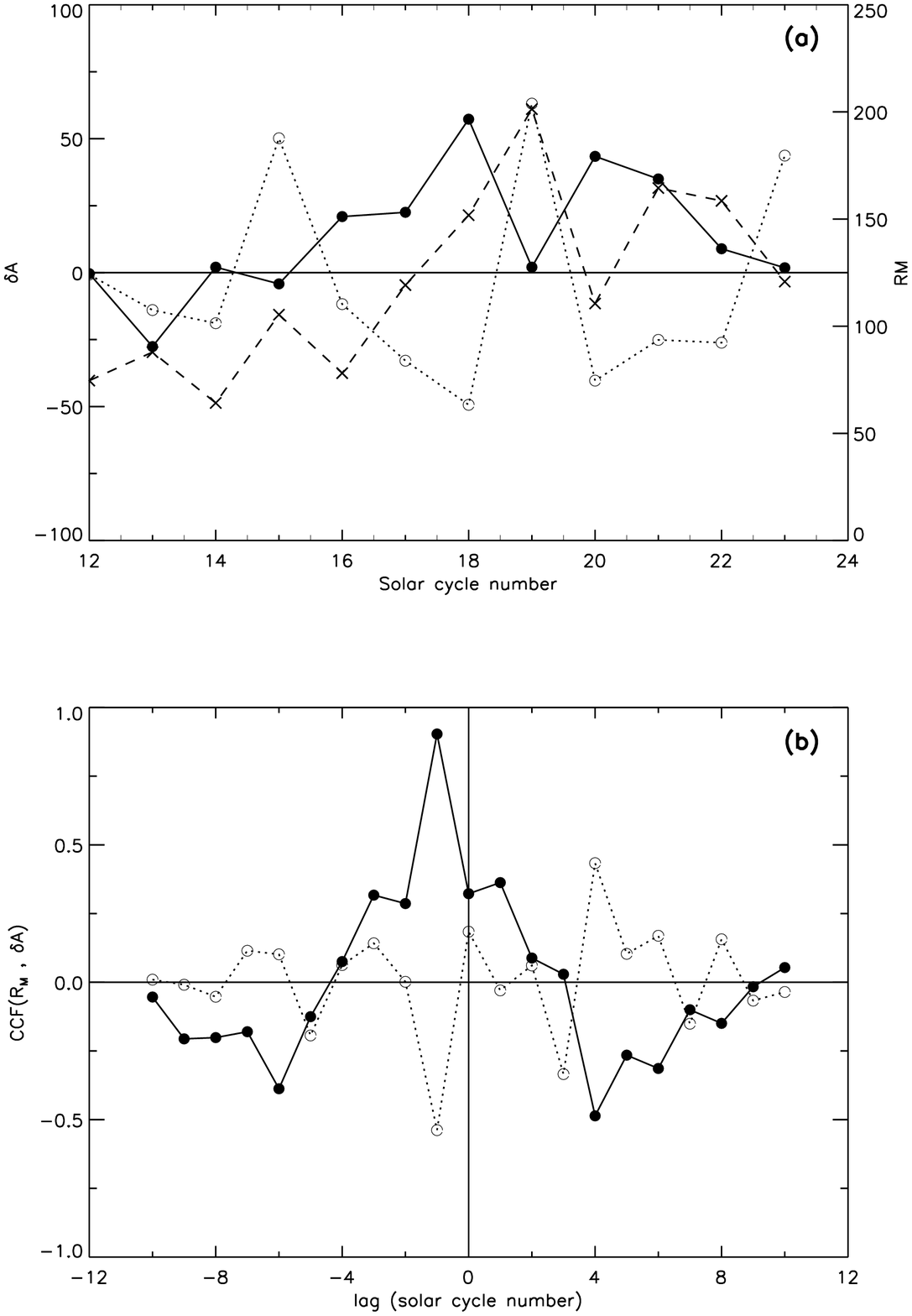}}
\caption{Plots of (a) the north-south differences (${\delta A}$),  $A_{\rm N}-A_{
\rm S}$ 
for the values of $A_{\rm N}$ and $A_{
\rm S}$  given in Table~1,  versus cycle number,
and (b) the corresponding CCF$(R_{\rm M}, {\delta A})$ versus lag,  
during the 
solar cycles~12\,--\,13 (Note:
a positive value of the lag
 indicates  that  
 $R_{\rm M}$ leads the ${\delta A}$).
The filled circle-solid curve and  open circle-dotted 
curve represent the ${\delta A}$ and the corresponding 
CCF$(R_{\rm M}, {\delta A})$  during $T_{\rm m}^*$ and $T_{\rm M}^*$, 
 respectively. In (a)  
the  cross-dashed curve represents the variation in $R_{\rm M}$.}
\end{figure}

\subsection{Variations in the North-South Asymmetries of the 
Area Sums}
Figure~2(a) 
 shows the variations in the north-south 
  differences ${\delta A} (T_{\rm m}^*) = A_{\rm N} (T_{\rm m}^*) - A_{
\rm S} (T_{\rm M}^*)$  
and ${\delta A} (T_{\rm M}^*) = A_{\rm N} (T_{\rm M}^*) - A_{
\rm S} (T_{\rm M}^*)$  
 during   the cycles 12\,--\,23,
 and Figure~2(b) shows the corresponding   cross-correlation 
functions, CCF$(R_{\rm M}, {\delta A})$, of the   
 cross-correlations between ${\delta A}$ and $R_{\rm M}$ (a positive value of 
lag indicates $R_{\rm M}$ leads ${\delta A}$).
In Figure~2(a) 
we have also shown the variation in $R_{\rm M}$ during cycles~12\,--\,23.   
In this figure it can be seen that there exits an approximate 
 anticorrelation ($r = -0.57$ ) between  
 ${\delta A} (T_{\rm m}^*)$ and ${\delta A} (T_{\rm M}^*)$. In the same figure 
it can also be seen that there is a strong suggestion 
of the existence of the $\approx$44-year periodicity 
in both ${\delta A}(T_{\rm m}^*)$ and ${\delta A}(T_{\rm M}^*)$. 
A 33\,--\,44 year periodicity is also found in   
the power spectrum analysis of the data on  
north-south asymmetries in  sunspot activity ~\cite{jg97,li02,boc05}
 and long-lived solar filaments~\cite{dd96}. 
This  periodicity is dominant in the north-south asymmetries 
of the solar
equatorial and the differential 
rotation rates
 determined from sunspot group data \cite{jg97},  
and it seems  to be present prominently  in 
the  climatically related  phenomena 
and in the Earth rotation rate, 
which may be related to the 44-year cycle 
 (double Hale cycle) of solar magnetic 
activity \cite{fh77,georg02}.

\subsection{The new and the Improved Predictions}

 In Figure~2(b) it can be seen that the 
cross-correlation function   CCF$(R_{\rm M}, {\delta A})$ of  $R_{\rm M}$ and ${\delta A} (T_{\rm m}^*)$ has  
 a strong peak (0.9) at ${\rm lag} = -1$,  
suggesting that ${\delta A} (T_{\rm m}^*)$  
 leads $R_{\rm M}$ by about 13 years, which is equal to the phase difference 
between  $A_{\rm N} (T_{\rm m}^*)$ and   
$R_{\rm M}$. Therefore, using  ${\delta A} (T_{\rm m}^*)$ of cycle $n$
 we can predict  $R_{\rm M}$     
of cycle $n+1$. The CCF$(R_{\rm M}, {\delta A})$ of $R_{\rm M}$ and  ${\delta A} (T_{\rm M}^*)$ has a 
maximum negative value also at ${\rm lag} = -1$, but its magnitude
 is inadequate for 
predicting $R_{{\rm M}, n+1}$ by using  ${\delta A}_n(T_{\rm M}^*)$.  

The correlation between ${\delta A}_n (T_{\rm m}^*)$ and $R_{{\rm M}, n+1}$  is  
 high ($r = 0.968$, corresponding  confidence level is $>99.99$)
  for cycles $n =$ 12 to 23. The  
corresponding  linear regression fit between 
 ${\delta A}_n (T_{\rm m}^*)$  and $R_{{\rm M}, n+1}$ is:
$$R_{{\rm M}, n+1} = (1.65 \pm 0.14)  {\delta A}_n (T_{\rm m}^*) + (99.8 \pm 3.9) , \eqno(3)$$
\noindent where uncertainties in the coefficients are
 the formal 1$\sigma$ (standard deviation) errors from the linear 
least-square fits. 

 Using  Equation~(3)     
the amplitudes  of the upcoming sunspot cycles can be predicted
  by about 13 year 
  advance. 
The result of the least-square fit 
is shown in Figure~3. 
 The  correlation 
between simulated  amplitude ($P_{\rm M}$) and  $R_{\rm M}$,  and 
also the level of the significance,  are found to be the same  
as that between  ${\delta A}_n$ and $R_{{\rm M}, n+1}$. 
Using 
 Equation~(3)  we obtained the value  
 $103 \pm 10$   for $R_{\rm M}$ of the  
  upcoming cycle~24. 

 It should be noted that since  $R_{\rm M}$ of  cycle~23    
is already known, hence it is included in the fittings of Equations~(1) and 
(3), although it is not needed for predicting  $R_{\rm M}$ of cycle~24. 
That is, using the values of $A_{\rm N} (T_{\rm m}^*)$ and ${\delta A}(T_{\rm m}^*)$ of cycle 23 and 
the linear-relationships determined from 
 the 9 pairs of data points correspond to the 
cycle pairs 12,13 to 21,22, we can get almost the same 
values for $R_{\rm M}$ of cycle~24 as found above.
Since  
cycle~23 will end soon, using  Equations~(1) and (3) it will be possible 
to predict an approximate value for  $R_{\rm M}$ of cycle~25 in about 3 years time. 
  $R_{\rm M}$ of cycle~24 is not needed for this.  

Since variation in  $A_{
\rm S} (T_{\rm m}^*)$ is weak (approximately  constant), 
hence the pattern of the variation 
of the corresponding ${\delta A} (T_{\rm m}^*)$ is almost the same as that
 of $A_{\rm N} (T_{\rm m}^*)$ ($r = 0.95$ between these quantities). Therefore, 
the  value obtained using Equation~(3) above for $R_{\rm M}$ of cycle ~24 
is approximately equal to the value obtained  
in Paper~I by using Equation~(1). 
However, the statistical significance of Equation~(3)  
 is  better 
 than that of Equation~(1).
 
\begin{figure}
\centerline{\includegraphics[width=12.0cm]{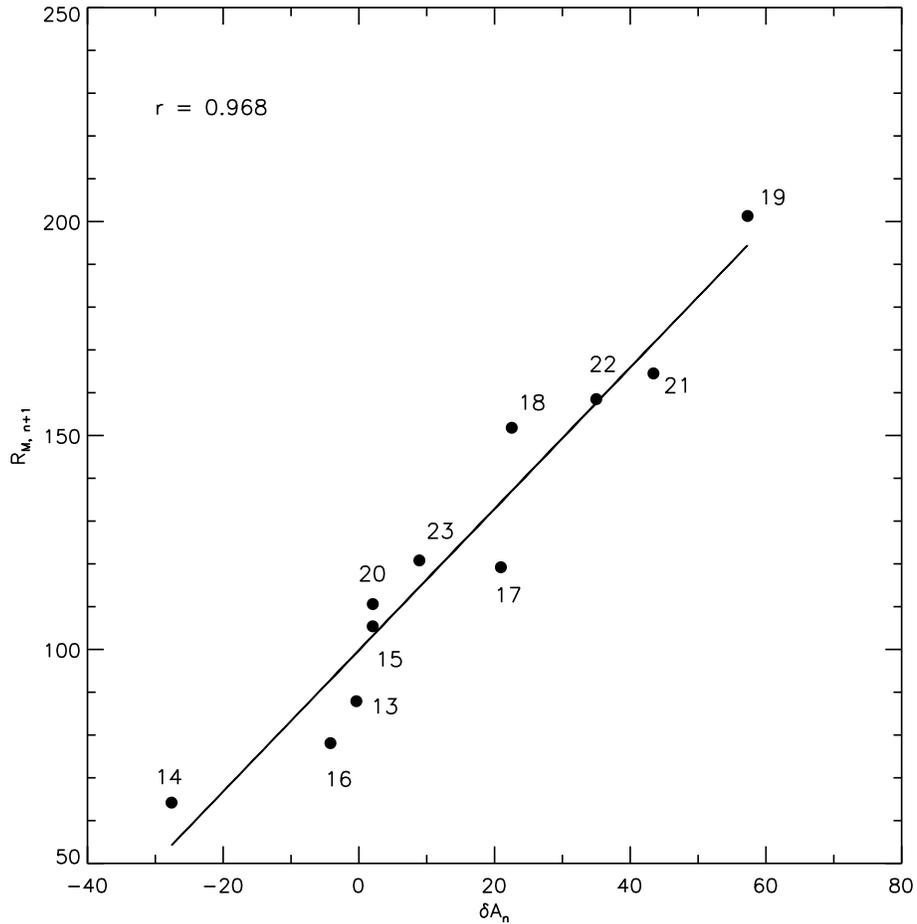}}
\caption{Plot of  the correlation between 
the north-south difference
 ${\delta A} (T_{\rm m}^*) = A_{\rm N}-A_{
\rm S}$ during $T_{\rm m}^*$ of
 cycle $n$ and $R_{\rm M}$ 
of cycle $n+1$, where  
$n = 12,...,22$ is the Waldmeier cycle number.   
  Near each data point the corresponding value of $n+1$ is shown.  
The solid line represents the corresponding linear relationship.   The 
value of the correlation coefficient ($r$) is also shown.
The  values of $A_{\rm N}$, $A_{
\rm S}$ and $R_{\rm M}$  
are the same which are 
 given in Table~1.}  
\end{figure}

\begin{figure}
\centerline{\includegraphics[width=12.0cm]{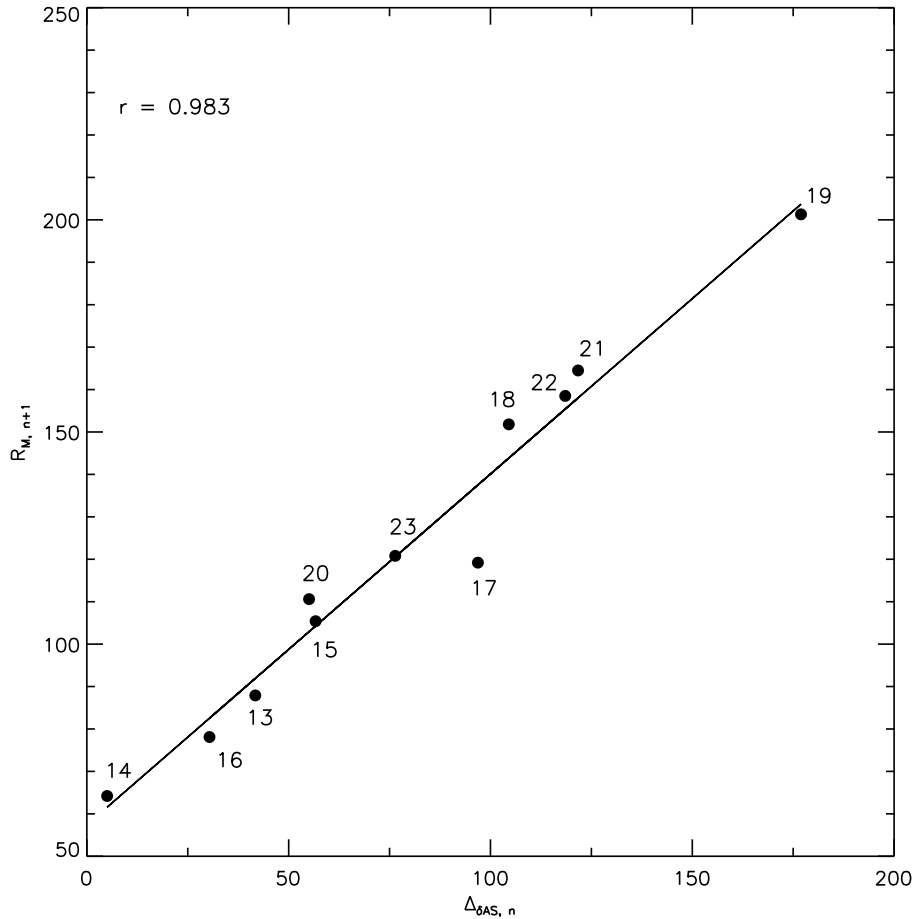}}
\caption{The same as Figure~3 but for 
the relationship between  $\Delta_{\delta{A{{\rm S}, n}}} = {\delta A}_n (T_{\rm m}^*) +A_{{\rm 
 S}, n} (T_{\rm M}^*)$ and 
$R_{{\rm M}, n+1}$.} 
\end{figure}

\begin{figure}
\centerline{\includegraphics[width=12.0cm]{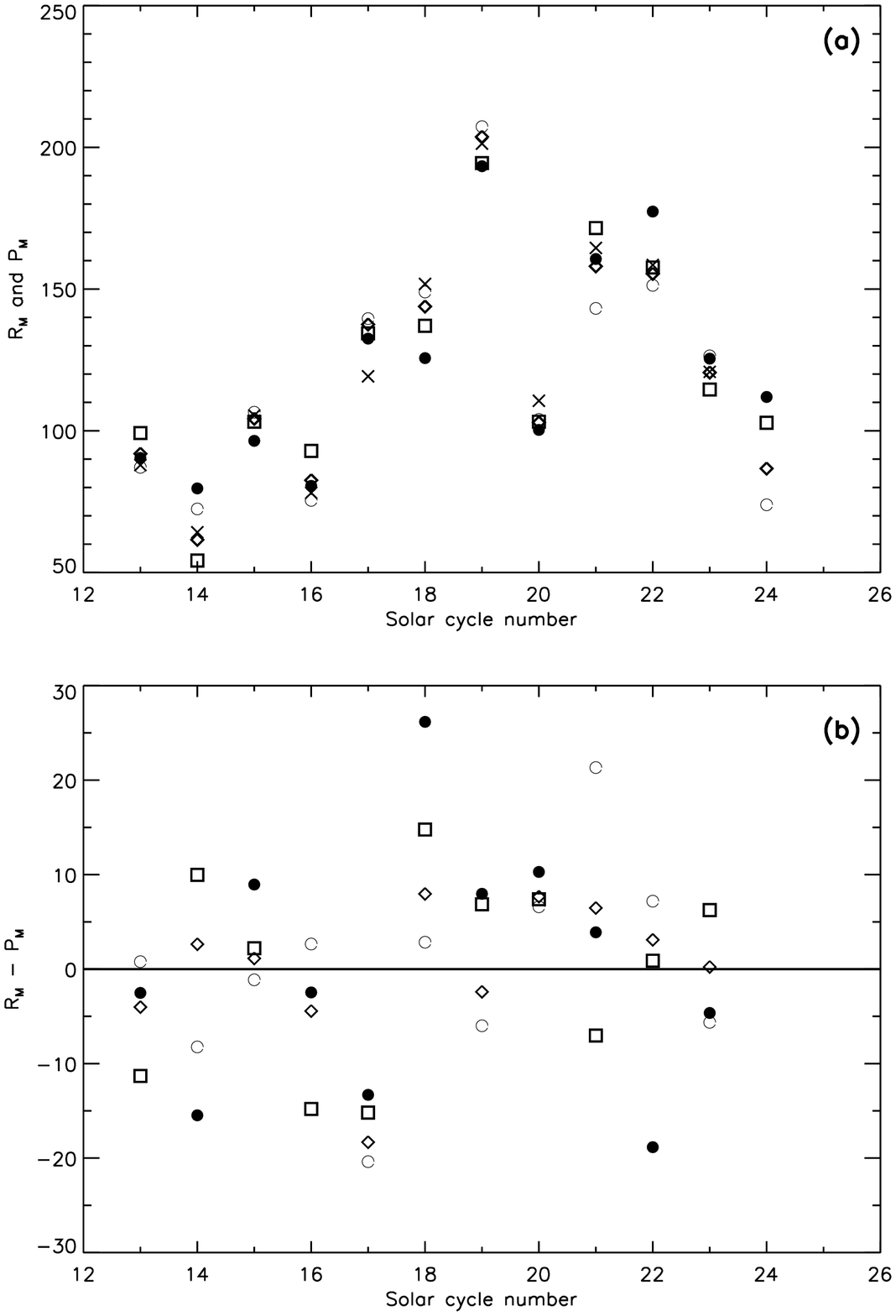}}
\caption{Plots of (a) the observed amplitudes ($R_{\rm M}$)  and the
simulated amplitude ($P_{\rm M}$) - simulated earlier using
 Equations~(1), (2) and
now using (3), (4) -
 and
(b) the differences, $R_{\rm M} - P_{\rm M}$,  against cycle number.
The cross represents
the $R_{\rm M}$.
 The  filled circle,  open circle,  square and  diamond
 represent the $P_{\rm M}$ and  the
difference, $R_{\rm M} - P_{\rm M}$, corresponding to Equations~(1), 2, (3) and (4),
 respectively.
In (a) the corresponding symbols at 
cycle~24 represent the values of 
 $P_{\rm M}$, 
 simulated using Equations (1), (2), and (3), and  represent the predicted
amplitude
of this upcoming cycle.}
\end{figure}

All the values, $112 \pm 13$, $74 \pm 10$ and $103 \pm 10$,   obtained 
 for $R_{\rm M}$ of  cycle~24 using Equation~(1), (2) and (3), respectively,  are 
 considerably (7\%, 39\% and 15\%, respectively) less than the value 
of $R_{\rm M}$ of cycle~23   
and consistent with the suggestion  that  
the level of activity is now  at the declining 
phase of  the current 
  Gleissberg cycle \cite{jbu05b}.

The  statistical significances of the value, $103\pm10$, predicted here
 by using Equation~(3)
 and the 
value, $74 \pm 10$,  
  predicted in Paper~I  by 
using Equation~(2) are equal,
 but the difference between these values 
is considerably large (it  is    
marginally less than   2$\sigma$). 
Therefore, 
we are unable to predict that  between these values which will become 
close to the real  $R_{\rm M}$ of cycle~24,   
although we  are confident about that the cycle~24 will be
 smaller than 
cycle~23.   

On the other hand,  Equations (1), (2) and (3) may represent the different 
components of  the total   
contributions of $A_{{\rm N}, n} (T_{\rm m}^*)$, ${\delta A}_n (T_{\rm m}^*)$ and $A_{{\rm 
 S}, n} (T_{\rm M}^*)$ 
to $R_{{\rm M}, n+1}$. Hence, it is necessary to find a 
 relationship between 
the combinations of  $A_{{\rm N}, n} (T_{\rm m}^*)$, ${\delta A}_n (T_{\rm m}^*)$ 
and $A_{{\rm 
 S}, n} (T_{\rm M}^*)$ and the $R_{{\rm M}, n+1}$.  We find that   
the $A_{{\rm NS}, n} = A_{{\rm N}, n} (T_{\rm m}^*) + A_{{\rm 
 S}, n} (T_{\rm M}^*)$, $i.e.$
 sum  of the area sums used in 
 (1) and (2),
 of cycle $n$   correlates very well ($r = 0.973)$ with the $R_{{\rm M}, n+1}$. 
Further,   the correlation between   
 $\Delta_{\delta{A{{\rm S}, n}}} = {\delta A}_n (T_{\rm m}^*) + A_{{\rm 
 S}, n} (T_{\rm M}^*)$ and $R_{{\rm M}, n+1}$
is very high ($r = 0.983$). From this 
relationship we get the following relation: 

$$R_{{\rm M}, n+1} = (0.83 \pm 0.05)  \Delta_{\delta{A{{\rm S}, n}}}  + (57.4 \pm 4.7) , \eqno(4)$$
\noindent whose  statistical significance is very high, $i.e.$, 
higher than those of all the relations (1)\,--\,(3).  Figure~(4) 
shows a scatter plot of    $\Delta_{\delta{A{{\rm S}, n}}}$ and $R_{{\rm M}, n+1}$.
By using Equation~(4) the prediction of 
 the amplitude of a cycle is  possible  
by about 9 years in advance with a high accuracy.
Using Equation~(4) we get $87 \pm 7$ for $R_{\rm M}$ of the upcoming cycle~24,
 which is about 
28\% less than the $R_{\rm M}$ of cycle~23. 

Figures~5(a) and 5(b)  show the plots of the 
simulated amplitudes ($P_{\rm M}$), simulated by using 
 Equations (1)\,--\,(4), 
 and 
 the corresponding differences, $R_{\rm M} - P_{\rm M}$,  
against solar cycle number 13\,--\,23. 
In Figure~5(a) we have also shown the variation in $R_{\rm M}$ during these cycles, 
and   
all the predicted values for 
$R_{\rm M}$ of cycle~24.  
The uncertainties in the predicted values     
 correspond to 1$\sigma$ 
values of  
 $R_{\rm M} - P_{\rm M}$ shown in Figure~5(b).
In  these figures it can be seen that the values of  
 $P_{\rm M}$  and  $R_{\rm M}$  closely agree each other,  in the  cases of  
 Equations (2)\,--\,(3). This agreement is much higher in the case 
of Equation (4).

\subsection{Will Cycle~25 be Stronger Than Cycle~24 ?}
According to the well-known Gnevyshev and Ohl, or G-O, rule \cite{go48}
 a preceding even numbered
 cycle is weaker than
its following odd numbered cycle. 
 However,  occasionally the G-O rule is violated. 
Figure~6 shows the annual variation of sunspot number during 1610\,--\,2006. 
As can be seen in this figure,  cycle~23 is an anomalous cycle in the sense
 that the cycle pair 22,23
 violated the G-O rule. 
The cycle~5 is also an anomalous cycle (the cycle pair 4,5 violated 
the G-O rule) and it is 
followed by 
the  week cycles 6 and 7.    
That is, a violation of the G-O rule seems to 
be followed by a few small cycles (see also~\opencite{jj05}). 
Therefore, the violation of the G-O rule by cycle pair 22,23 
also indicates that the next cycle~24 (probably even the cycle~25)  
will be a weak cycle. 

In principle the G-O rule can be used  for predicting $R_{\rm M}$ of 
an odd numbered cycle from that of its preceding even numbered cycle. 
But this is  possible only when we  know   
 in advance that the  even- and odd-numbered
 cycle pair will not violate the G-O rule.
 For instance, in order to predict  $R_{\rm M}$  
of cycle~25 from that of cycle~24 (when it will be known)
 by applying  the G-O rule, 
it is necessary and essential 
 to know in  advance that cycle pair~24,25  will not violate the G-O rule.
Recently, \inlinecite{jj05} shown that it may be possible 
to forecast the epochs of the violations of the G-O rule 
  well in advance  from the epochs of retrograde  motion of the Sun 
about the solar system barycenter. 
Except that so far  no  plausible method is available 
for predicting the violation of the G-O rule. 
As can be seen in Figures~1(a) and 1(b)   
 the values of $A_{\rm N} (T_{\rm m}^*)$ and  $A_{
\rm S} (T_{\rm M}^*)$ are higher in an even numbered
 cycle than in  
its following odd numbered cycle, say 'inverse G-O rule',  
except in case of cycle pair  20,21.
   That is, this inverse G-O rule  
 in  $A_{\rm N} (T_{\rm m}^*)$ and    $A_{
\rm S} (T_{\rm M}^*)$
 is violated by about 22 years (a Hale cycle) 
earlier 
than the corresponding violation of the G-O rule in $R_{\rm M}$.  
Hence,   the studies  of variations
 in  $A_{\rm N} (T_{\rm m}^*)$ and    $A_{
\rm S} (T_{\rm M}^*)$ may also help for predicting a future even- and 
odd-numbered cycle pair which will violate 
the G-O rule, 
 by about 22 years (a Hale cycle) 
advance.  Since cycle pair~22,23 satisfies the aforementioned inverse G-O rule, 
hence, probably cycle pair 24,25 will satisfy the G-O rule, $i.e.$ probably  
cycle~25 will be stronger than cycle~24.

In Figures~1(a), 1(b) and 2(a) it can  be seen that
 a  $\approx$44-year periodicity  
exists  
  in $R_{\rm M}$ also. However,    
it is not as strong as the corresponding periodicity in  $A_{\rm N} (T_{\rm m}^*)$ and 
 $A_{
\rm S} (T_{\rm M}^*)$ and in their corresponding north-south differences.  
\inlinecite{roz94} found that in sunspot data  
 only 11-year periodicity is statistically 
significant and all the remaining  periodicities 
are only minor fluctuations. 
As can be seen in Figure~2(a) the $T_{\rm m}^*$ epochs of the cycles 15, 19, and 23 are at the  minimum epochs 
 of the 
$\approx$44-year cycle in ${\delta A}(T_{\rm m}^*)$, whereas the corresponding epochs of 
$T_{\rm M}^*$ are at the maximum epochs of the  44-year cycles  in ${\delta A} (T_{\rm M}^*)$. 
In Figure~6 it can be seen that  
the 
 odd numbered cycles 11, 15 and 19  are followed by the 
weak even numbered 
cycles 12, 16 and 20.  
Cycle 7 is also weak  and  followed by the strong cycle~8. However,  
the cycle    
 pair 8,9 violated the G-O rule. As pointed out  above  probably 
 the cycle pair 24,25 will not violate 
 the G-O rule. All these indicate that 
cycle 24 will be 
  weaker than cycle~23. 
Thus,   from the $\approx$44-year periodic variations  
of ${\delta A}(T_{\rm m}^*)$ and 
${\delta A} (T_{\rm M}^*)$  one can infer  that the upcoming cycle~24 will be 
weaker than cycle~23. 

 From the patterns of  ${\delta A} (T_{\rm m}^*)$ and 
${\delta A} (T_{\rm M}^*)$ we can also infer that the upcoming cycles~24  and 28 may be 
 at the beginning ($i.e.$, the end of the current Gleissberg cycle) 
and the ending  minimum epochs of the next 
44-year cycle, respectively (the cycles 32, 36 and so on 
may be also weak cycles). 
 In  both ${\delta A} (T_{\rm m}^*)$ and ${\delta A} (T_{\rm M}^*)$, and also in $R_{\rm M}$,  the patterns of the   
44-year cycles are considerably different from one 44-year cycle to another. 
In case of ${\delta A} (T_{\rm M}^*)$,  the  pattern of its variation 
  seems to be changing in the alternate 80\,--\,90-year cycles, indicating 
 that the pattern of the next 44-year cycle  may be the same 
as that of the current 44-year cycle. Therefore, since the cycle~21 is
 stronger 
than cycle~20,  we can expect that the cycle~25 will be stronger than
 cycle~24. This is consistent with the speculation above, $viz.$        
probably the cycle pair 24,25 will satisfy the G-O rule.

\begin{figure}
\centerline{\includegraphics[width=12.0cm]{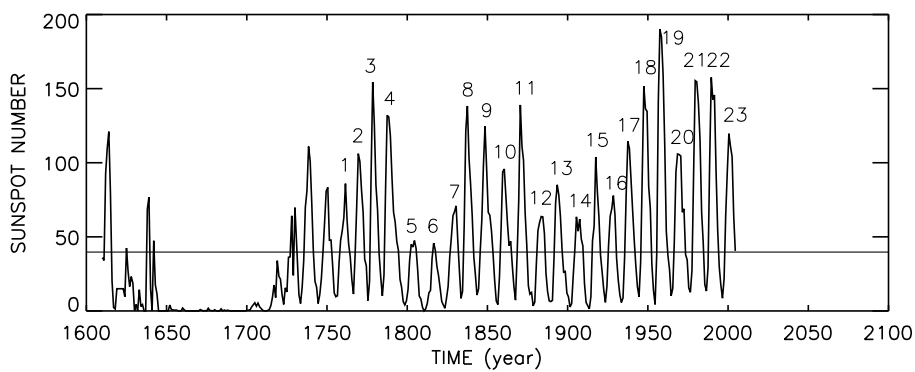}}
\caption{Plots of annual sunspot number against time. Near the peaks of 
the cycles corresponding Waldmeier solar cycle numbers are shown.}  
\end{figure}

\section{Discussion}

\subsection{A Physical Interpretation of the Relationships Found Above}
According to Equations (1)\,--\,(4),  
a large/less value of  any of  $A_{\rm N} (T_{\rm m}^*)$, $A_{
\rm S} (T_{\rm M}^*)$, 
${\delta A} (T_{\rm m}^*)$ and  $\Delta_{\delta{A{\rm S}}}$, $i.e.$ 
a large/less emergence 
of the magnetic flux (unsigned)  
in  the corresponding $0^\circ-10^\circ$ latitude 
interval of the northern or the southern hemisphere  during   
  $T_{\rm m}^*$ and $T_{\rm M}^*$,  of a cycle $n$ implies
a  large/less emergence of the magnetic flux at $T_{\rm M}$ 
of the next cycle $n+1$. 
It should be noted here that the epochs $T_{\rm M}^*$ of cycle $n$ and $T_{\rm M}$ 
of cycle $n+1$ are 
around the begin and the end of a same polarity half solar magnetic cycle which 
comprises the declining phase of the cycle $n$ and the rising phase  of 
the cycle $n+1$,  
whereas
the epoch $T_{\rm m}^*$ is around the middle of the opposite polarity half solar 
magnetic cycle   
 which comprises the declining phase  of the 
cycle  $n-1$ and the rising phase of the cycle $n$. 
There exist strong correlations between $A_{{\rm N}, n} (T_{\rm m}^*)$,  
${\delta A}_n(T_{\rm m}^*)$ and $A_{{\rm 
 S}, n} (T_{\rm M}^*)$, 
whereas there is no significant correlation either between 
$A_{{\rm N}, n+1} (T_{\rm m}^*)$ and 
$A_{{\rm 
 S}, n} (T_{\rm M}^*)$ or between ${\delta A}_{n+1} (T_{\rm m}^*)$ and $A_{{\rm 
 S}, n} (T_{\rm M}^*)$. 
This confirms  that
 $A_{\rm N} (T_{\rm m}^*)$ 
and ${\delta A} (T_{\rm m}^*)$ lead $A_{\rm 
 S} (T_{\rm M}^*)$ by about 5 years. 
It also suggests that  the 
emergence/cancellation of the magnetic flux at  $T_{\rm m}^*$ of cycle $n$ 
  may  influence 
on the emergence/cancellations of the magnetic flux at both the 
$T_{\rm M}^*$ of cycle $n$ 
and $T_{\rm M}$ of cycle $n+1$.  The emergence/cancellation of the flux at  
$T_{\rm M}$ of cycle $n$ may not
 have an influence on the emergence/cancellation of the flux at $T_{\rm m}^*$
 of cycle $n+1$. 
 Even 
it may not have an influence on the emergence of the flux at $T_{\rm M}$ of cycle
 $n+1$.
That is, the high correlation
between $A_{{\rm 
 S}, n} (T_{\rm M}^*)$  and $R_{{\rm M}, n+1}$ may be
due to the influence of the emergence/cancellation of the magnetic flux  
at $T_{\rm m}^*$ of cycle $n$ on the emergence/cancellation of the flux  
at both
epochs $T_{\rm M}^*$ of cycle $n$ and   $T_{\rm M}$ of cycle $n+1$.
The existence of an  'inverse G-O rule' 
in  $A_{\rm N} (T_{\rm m}^*)$ and $A_{\rm 
 S} (T_{\rm M}^*)$ was pointed out in Section~2.5. 
It may be interesting to note here that a similar behavior was also 
 found 
in the   latitudinal gradient of the solar rotation 
determined from the sunspot group data~\cite{jbu05a}.  
Hence, the  emergence/cancellation of the magnetic flux 
 at the  epochs $T_{\rm m}^*$ and $T_{\rm M}^*$ of cycle $n$ 
and $T_{\rm M}$ of cycle $n+1$, i.e. all the relationships found above, 
 may be related to the   
long-term variations, $\approx$ 22 years and longer, in the differential 
rotation~\cite{jj03,jbu05a,jbu05b}.

\subsection{The Role of the Solar Meridional Flows for the Relationships Found
 Above} 
It is well accepted that the solar dynamo, which seems to be 
located  near the base of the Sun's convection
zone,   generates  the solar 
magnetic field for solar activity and the  solar  cycle~\cite{rw92,os03}.
The Sun's polar fields, solar meridional flows and differential rotation  
are important ingredients in the dynamo 
models (\opencite{bab61},\inlinecite{ub05} and references therein). 
The solar meridional flows can transport magnetic flux 
and play a major role for the  
 magnetic flux cancellation and the polar field 
reversals (\inlinecite{wang04} and references therein). 
As  already suggested in Paper~I,
the relationships between  $A_{{\rm N}, n} (T_{\rm m}^*)$ and $A_{{\rm 
 S}, n} (T_{\rm M}^*)$ and $R_{{\rm M}, n+1}$ 
 may have a physical  relationship with 
the solar magnetic cycle
 and the temporal variations of the 
 solar equatorial rotation rate and the meridional flows. 
Figure~7 shows the variations in the mean 
meridional velocity, $v'$,  of sunspot groups during the  
odd and the even numbered solar cycles determined by superposing the spot group 
data during cycles 11\,--\,23 according to the years relative to the 
nearest sunspot minimum epochs 1867, 1879, 1890, 1902, 1913, 1923, 1934, 
1944, 1954,1965, 1976, 1987, and 1997.
 The data reduction and analysis are the same as in \inlinecite{ju06}. 
Now we have also used the spot group data during cycles~22\,--\,23.
This  increased the size of the data considerably. Hence,  
it enabled us to  determine somewhat  reliable   
variations in the mean meridional motions of spot groups in the 
northern and the southern 
hemispheres during the odd cycles (11, 13, 15, 17, 19, 21, 23) 
 and the even cycles (12, 14, 16, 18, 20, 22).     
\inlinecite{tuo52} noticed the existence of a difference in the 
meridional motions of spot groups during the odd and the even cycles 
 and suggested the existence of a  22 year periodicity in the solar 
meridional flow. 
 \inlinecite{pid76} assumed that there exist 22-year meridional 
 oscillations which  change the sign of the angular 
velocity gradient with respect to the relic magnetic field lines causing 
22-year solar magnetic oscillations. 
 Hence, in order to look for 
a longer than 11-year periodicity ($\approx$22 year periodicity)
 in the  meridional flow,
in Figure~7 the values of the mean meridional motion of spot groups  
during the odd cycles are 
plotted against the years 1 to 12  and  the corresponding 
values of the even cycles are plotted against  
the years 13\,--\,23. 
 A vertical dashed line is drawn at the year 12.5    to 
identify the ends of the
odd numbered cycles and the beginnings of the even numbered cycles.   
This figure reveals   more  information 
than the Figure~2 of \inlinecite{ju06}, where
the data from both the odd- and  even-numbered cycles  are  combined.

As can be seen in Figure~7 
there are 
  considerable differences in the mean meridional motions of the spot 
groups in a given 
hemisphere  
during  the odd  and the even cycles.
For example,   at the beginning of an odd cycle 
the motions are equator-ward directions in the  
$0^\circ - 10^\circ$ latitude intervals of both the  northern and 
the southern hemispheres 
(the motions in the northern and the southern hemisphere are southbound 
and northbound, respectively). But  
this is opposite at  the beginning of an even cycle, $i.e.$ 
the motions are in pole-ward directions in $0^\circ - 10^\circ$ latitude 
intervals of both the hemispheres. There is an indication that  
the motion in the $0^\circ-10^\circ$ latitude 
interval of the northern hemisphere rapidly changed (in about two years time)
 from the pole-ward (northbound) direction 
  at the end of an even cycle, say $n$,   to the equator-ward (southbound)
at the begin of the following odd cycle, say $n+1$.  
In  $0^\circ - 10^\circ$ latitude interval   
of the southern hemisphere 
the motion is not significantly different from zero at the end of 
an even cycle $n$, but it became significantly different from zero and 
equator-ward    
(northbound) direction 
 at the begin of the following odd cycle, $n+1$.
These opposite directions 
   of the mean meridional motions of the spot groups  
  suggest that around the begin of an odd numbered cycle the solar  
meridional flows may transport  magnetic flux across the equator  
 causing   cancellation of a large amount  
of the magnetic flux around the equator. 
 The equator-ward meridional flows may include  large contributions 
of the down flows at the active regions in the sunspot latitude belt. 
The cancellation of the flux      
 is relatively weak or even absent around the begin of an even numbered cycle, 
because the motions are pole-ward in  $0^\circ - 10^\circ$ 
latitude intervals of both the hemispheres.
 During the declining  phase of any cycle,  
 until approximately 
one year before the end of the cycle 
 the motions are equator-ward directions   
in $0^\circ-10^\circ$  latitude intervals
of both the northern and the southern hemispheres. 
This may  cause cancellation of some amount of magnetic flux  around
 the equator during  the declining phases of the cycles.
As we have already found in Paper~I, the time-intervals
 $T_{\rm m}^*$ and $T_{\rm M}^*$  included the epochs when the  
motion is changed to pole-ward to equator-ward.

 All the aforementioned suggestions of the variations in 
the mean meridional motions of the spot groups 
are very well consistent with the 
differences between the odd and the even numbered  
cycles in the  values  of  
 $A_{{\rm N}, n} (T_{\rm m}^*)$, 
${\delta A}_n (T_{\rm m}^*)$ and $A_{{\rm 
 S}, n} (T_{\rm M}^*)$ and the physical interpretations  
 drawn in Section 3.1, above.
In addition, the overall pattern of the mean meridional motion of the spot
groups 
in $0^\circ -10^\circ$ latitude  interval of the  
southern hemisphere suggests that 
  about 9\,--\,10 year periodicity is relatively dominant 
 in the solar meridional flow 
 in this latitude interval. 
 The corresponding pattern in the $0^\circ - 10^\circ$ 
latitude interval of  
 the northern hemisphere suggests that 
 a relatively 
longer periodicity of about 15\,--\,20 years (probably even longer than this, 
i.e. $\approx$30 years),  is dominant 
in the solar meridional flow in this latitude interval. 

Here we  concentrated  mainly on the mean solar cycle variation  
of the meridional motions of the spot groups in  $0^\circ-10^\circ$ 
latitude intervals of the northern and the southern hemispheres.
However, as can be seen in Figure~7 the behavior of the mean motions 
in the high 
latitudes are substantially  
 different from those of the low latitudes. It is    
consistent with the discussion in Paper~I on the implications of 
  the  north-south asymmetries in the latitude distributions of  
the solar flare activity  during a  solar cycle ($e.g.,$~\opencite{garcia90}), 
 and on the large north-south asymmetry in the sunspot activity 
during the later Maunder minimum (see \opencite{sokn94}).

It is interesting to note here  
that during cycle~23 the mean meridional motion of 
the sunspot groups is stronger 
than  
that during about last 9\,--\,10 cycles. That is,  
during cycle~23 (around maximum epoch)  the motion is strongly pole-ward 
in the northern hemisphere and strongly equator-ward in the southern hemisphere
 (see Figures~8 and 9 in~\opencite{ju06}). 
That is, the  overall mean motion is northbound and may be responsible 
for the cancellation of a large amount of 
magnetic flux during this cycle (mainly during the declining phase)
and the flux cancellation might be
 relatively more in the northern
 hemisphere than in southern hemisphere. 
Hence, the cycle~23 is weak and the  
 activity is slightly more 
 in the southern hemisphere than in the 
northern hemisphere, mainly during the declining phase of this cycle.

\begin{figure}
\centerline{\includegraphics[width=12.0cm]{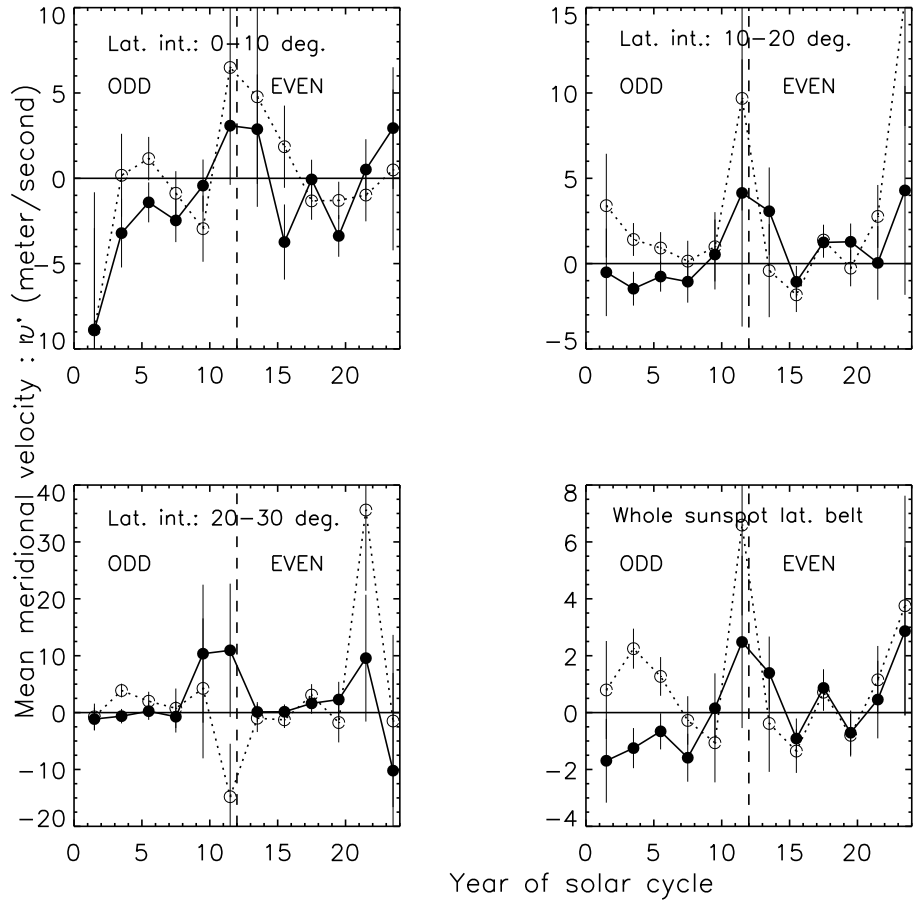}}
\caption{Variations in the mean meridional velocity, $v'$, of the sunspot groups
in the northern hemisphere (filled circles and solid curves)
 and the southern hemisphere
(open circles and dotted curves)
during the odd numbered solar cycles (11, 13, 15, 17, 19, 21, 23) and the 
even numbered cycles (12, 14, 16, 18, 20, 22). 
The minima of the odd- and even-numbered cycles correspond to
 year one and thirteen, respectively.  Averages are 
taken over 2-year intervals.  
The vertical dashed line  
at year 12.5 is drawn to identify the ends of the odd numbered cycles
 and the beginnings of the 
even numbered cycles.  In both the northern and southern hemispheres,  
the positive values of the velocity indicate pole-ward motions and the negative
 values indicate
 equator-ward motions.  
The horizontal solid lines represent the zero values of $v'$.}
\end{figure}

\subsection{Comparison of our Prediction with the predictions Based on the Dynamo Models}
A large number of forecasting methods 
(precursors, spectral analysis, non-linear dynamics, 
solar system dynamics, etc.)  are used to predict the amplitudes of
the solar cycles. There have been many predictions for the $R_{\rm M}$ 
of cycle~24,
but their range is very wide (for the list of these predictions
see~\inlinecite{kane07a} and \inlinecite{obrid08}). 
Therefore, the  solar cycle~24  prediction panel  found  difficult 
to came out consensus on any  single value and   
  supports to the
two possible peak amplitudes, $90 \pm 10$ and $140 \pm 20$,
 for the smoothed monthly value of the  
international sunspot number of this cycle 
 ({\tt http://www.sec.noaa.gov/Solarcycle/\break sc24/statement\_01.HTML}).
 Our prediction is on the  low value side.

Among all the methods of forecasting of the maximum amplitudes of 
the  solar cycles, 
precursor method based on the  correlation between the strength 
of a geophysical phenomenon in the declining phase of a cycle and the 
strength of its immediate following cycle,    
 initiated by~\inlinecite{ohl66}, seems to be most 
successful one~\cite{kane07a}. 
Recently, \inlinecite{hw06} and   \inlinecite{kane07a} by 
using the current strength of the geomagnetic   
 $aa$  index  as a precursor predicted  
$ 160 \pm 25$ and  
$142 \pm 24$, respectively, for $R_{\rm M}$ of cycle~24. 
 \inlinecite{sch78} used  for the first time  the strength of the polar magnetic
 field at the preceding 
minimum epoch of  a cycle as the precursive indicator for the
 amplitude  of the 
same cycle. This method has a physical background in the sense that 
   the strength of the polar field
 during the minimum of a cycle is an important 
ingredient for dynamo models of the solar cycle. 
Using this method and the current weak strength of 
the polar field near the end    
of the ongoing cycle~23, measured in Wilcox Observatory,
  \inlinecite{sch05} predicted $80 \pm 30$  
and \inlinecite{sva05} predicted $75 \pm 8$  for $R_{\rm M}$ of cycle~24.
Recently, 
on the basis of the  
current strength of the polar field 
measured in the Mount Wilson 
Observatory, which is weaker than the strength of the polar filed  
at the preceding minimum of cycle~22,
 Ulrich, R.K. and  Boyden, J.E. also predicted that cycle~24 will be weaker
 than cycle~23   
 (private communication from Ulrich, R.K. and  Boyden, J.E., 
 {\tt http://astro.ucla.edu/\~\ ulrich/}). 
\inlinecite{hw04} suggested a large value $145 \pm 30$ for $R_{\rm M}$ of cycle~24,  
 using  the fast equatorial drift rate of sunspot activity 
($i.e.$ fast equatorial meridional motion)  during 
the declining phase of cycle~22 found by~\inlinecite{hnwr03}. 

\inlinecite{ddg06} and \inlinecite{dg06} by inputting
the data on  sunspot group area during the cycles~12\,--\,23
to a flux-transport dynamo model,  predicted a very 
large amplitude, 150\,--\,180, for $R_{\rm M}$ of cycle~24.
 One of the main assumptions
of their model is  a long, about 17\,--\,23 year, magnetic memory (a slow
diffusion of magnetic field).
 On the other hand \inlinecite{ccj07} and \inlinecite{jcc07} by inputting
the strength of the observed polar field, $i.e.$, by using the same data 
that used by~\inlinecite{sva05},  
into  another type of flux-transport model, 
theoretically confirmed the  low values
predicted by~\inlinecite{sch05} and \inlinecite{sva05}.
This model assumes a short magnetic memory of 
about 5-year time (a fast diffusion of magnetic fields).
\inlinecite{cs07} studied the origin of the predictive skill of 
the  flux-transport dynamo models,  theoretically and observationally
 by using  
both the sunspot area data during cycles 12\,--\,23 and the sunspot numbers 
 during cycles 1\,--\,23, and found that the assumed values for  
the parameters of the 
aforementioned  flux-transport
dynamo models are ineffective, so that  
the predictive power lies in    
the temporal behavior of the observational  data used in these 
 models.  
They have concluded that  
since the stronger cycles tend to rise faster to their 
maximum activity (Waldmeier effect),   
the origin of the predictive power in the  above mentioned 
flux-transport dynamo
 models is    
 the shifts of the minimum epochs of the cycles due to 
temporal overlapping of the cycles, and hence  
the  predictive skill does not  require a magnetic memory.
However,  \inlinecite{ddg06} and \inlinecite{dg06}  assumed  a
 long, 17\,--\,23 year, magnetic memory.  
Hence, a major contribution to their prediction for cycle~24  
 might have also come before the overlapping period of the 
 cycles~23 and 24. In addition, recently \inlinecite{dgd08} found that 
the Waldmeier effect is not present in sunspot area.
  In  our case,  
 first of all we have used the data on the spot groups in 
$0^\circ-10^\circ$
 latitude intervals only.  
Obviously, the epoch $T_{\rm m}^*$ of a cycle  $n$ is quite  far away from 
the overlapping 
period of the cycles $n$  and $n+1$.
The  $T_{\rm M}^*$ of a cycle $n$ is just one year away from the 
$T_{\rm M}$ of cycle $n$, $i.e.$, 
it is also considerably far away from the 
time of the overlapping of the cycles $n$ and $n+1$. 
As already found  in Paper~I,   $T_{\rm m}^*$  and  $T_{\rm M}^*$ of a cycle $n$ 
 are close to the epochs  when the polarity reversals of the global 
magnetic fields   take place~(\opencite{mak01}, 2003).
 Of course, as already mentioned above the $T_{\rm M}^*$ of cycle $n$  
and the $T_{\rm M}$ of cycle $n+1$
are in the same  polarity half magnetic cycle.

The values we have predicted  for $R_{\rm M}$ of cycle~24,
 by using Equations~(2) and (4), are  close to the low values
predicted by~\inlinecite{sch05}  and \inlinecite{sva05}. 
However, by using  
our method the prediction can be made by  at least four years earlier than
their method. 
 Both \inlinecite{ddg06}, \inlinecite{dg06}  and ourselves
used the same spot group data which are taken from David Hathaway's website
  (we have used the raw data files which contain the daily measurements).  
They have obtained a high correlation ($r = 0.98$) 
between the simulated and the observed strengths of 
the eight cycles, 16\,--\,23.  Our prediction 
is much more statistically significant than their prediction, 
because we have $r =0.98$ from eleven  data points.  
Since their model needs 17\,--\,23 year  magnetic memory, hence,  in 
 their simulation for a cycle $n+1$ a major contribution might  have come 
also from cycle   
 $n-1$ and even from before it, besides  a contribution from the 
 cycle  $n$. 
Their predicted  value for $R_{\rm M}$ of  
 cycle~24 is large probably  because of 
a large contribution to it came from  the  large cycle~22 and   even  
also from  cycle~21 which is also a large cycle, 
besides a contribution 
by cycle~23.
They have  used a constant value,   
 14 m s$^{-1}$, for the speed of the meridional flow, whereas the speed  
of the meridional flow may  
vary considerably ($cf.$, Section~3.2 above and~\opencite{ju06}). 
In addition, 
 they have assumed a profile for  cycle~24  
whose size is almost 
equal  to that of cycle~23. However, they  found a large value even without 
assuming any profile for cycle~24 and were able to  
correctly  reproduce the small amplitude of cycle~20.  
Our case is also based on the long-term variations in 
the sunspot activity. But 
for the prediction of a cycle $n+1$
 the main  contribution 
  comes from around the preceding minimum and near the maximum of cycle $n$ 
(for predicted $R_{\rm M}$ of  
cycle~24 the main contribution comes from  cycle~23)
 and that too 
only from the  $0^\circ - 10^\circ$ latitude interval. 
However,
 our case does not rule out  
the possibility of some minor  contributions of cycle $n-1$  and  
around the minimum of cycle $n+1$ itself, particularly  from the
 high latitudes. In fact, it may be necessary to take into account of such
 contributions,   
 because the magnetic fields at different latitudes 
 during different time-intervals of  a previous  
 cycle might contribute to the activity at the same or different 
latitudes during  the next cycle.  
If we can include such contributions, 
then we may get an improvement in the corresponding  correlation  
of Equation~(4),  from 98\%  to 100\%,  and  a value for $R_{\rm M}$ 
of cycle~24     
which may be considerably different  than the values predicted here.  
Recently, \inlinecite{kane07b} found a reasonably good 
correlation ($r = 0.89$)  
between the sum of the sunspot group numbers in the latitude intervals 
$10^\circ-20^\circ$ and $20^\circ-30^\circ$ 
of the northern hemisphere during a preceding minimum of cycle $n$ 
and $R_{\rm M}$ of cycle $n+1$, 
 and 
predicted 129.7$\pm$16.3 for $R_{\rm M}$ of cycle 24. 
As we have already mentioned in   
Section~2.4,  all our predictions are  consistent with   
that now the level of activity is at the declining phase of the current 
Gleissberg cycle~\cite{jbu05b}. However,  the earlier result itself is 
 yet to be 
confirmed.

\subsection{Comparison of our Prediction with the Predictions Based on the 
Spectral Analyses and Magnetic Oscillations Models} 
The  properties of the 
solar cycles can also be  explained on the basis of superpositions 
of the  dominant modes 
of the Sun's global magnetic
 oscillations
 whose frequencies  
 equal to that of  the  solar magnetic cycle and a 
few harmonics of it ($e.g.,$~\opencite{bra88}; \opencite{sten88}; \opencite{gj90}; \opencite{gj95}; \opencite{juck03}). Such a model 
 has a strong predictive power ($e.g.,$~\opencite{gj95}).  
A number of authors used various 
spectroscopic methods  and predicted around 100 
for $R_{\rm M}$ of cycle~24 ($e.g.,$~\opencite{kane99}; \opencite{echer04}).  
Recently, \inlinecite{kmh06} modeled the solar cycle
as a forced and damped harmonic oscillator.  \inlinecite{kmh08}  by inputting 
the amplitudes,  
frequencies, and phases of 22 cycles (1755\,--\,1996) derived from the 
aforementioned  model 
into an autoregressive model,  
predicted the periods and the amplitudes of the next 
fifteen solar cycles. 
 His prediction for $R_{\rm M}$ of cycle~24 is $110\pm11$. 
It may be interesting to note here that some cycles, including cycle~23,  contain double peaks.
\inlinecite{gn67} interpreted this property of the solar cycles  
as  an 11 year cycle consisting of
 two processes (waves ?)
 with different physical properties and the shape 
of the cycle depends on the way these processes overlap.
That is, the occurrence of a double peak  cycle   
 may be mainly related to the difference  
 between the phases of the 
processes (waves) involved. That is, a large difference 
in the initial phases of the waves may be responsible for  a small 
and broad or double  peak, and also long period, cycle. 
A number of authors also explored the possibility that  
the solar cycle is a sum of two periodic 
functions ($e.g.,$~\opencite{merz97}). 
The variations in ${\delta A} (T_{\rm m}^*)$ and $A_{
\rm S} (T_{\rm M}^*)$ may represent
 the two
 dominant magnetic waves (quadrupole and dipole components of the 
 global magnetic oscillations) in the Sun, 
 whose superposition may  
be responsible for the long-term variations in 
the amplitudes of solar cycles. The $A_{\rm N} (T_{\rm m}^*)$ and  
 ${\delta A} (T_{\rm m}^*)$  lead $A_{
\rm S} (T_{\rm M}^*)$  by about 4 years. 
The difference between $T_{\rm m}^*$ of  cycle $n$ and $T_{\rm M}$ of  
 cycle $n+1$ is 14\,--\,18 years, and that  
between   $T_{\rm M}^*$ of cycle $n$ and $T_{\rm M}$ of cycle $n+1$ is 9\,--\,12 years,
 which may represent the periods of the aforementioned 
waves. Such  periodicities are found to exist also in 
the solar differential rotation determined from sunspot group 
data~\cite{jg95,jk99,jj03,jj05,brw06}. The existence of $\approx$ 22-year
periodicity is detected also in the solar rotation determined from 
 the data on large-scale magnetic fields measured using the 
 H$\alpha$ line, magnetographic observations, and spectral-corona 
observations~\cite{tlatov07}. We think 
the combined effect of the Sun's rotation and the inclination of the 
Sun's equator 
to the ecliptic may have a major role in all the  
 relationships found above.

\section{Summary and Conclusions}
Using the values of the sums of the areas, $A_{\rm N}$ and $A_{
\rm S}$,  of the 
spot groups in 
$0^\circ-10^\circ$  latitude intervals of the northern and the southern hemispheres and  
during  the time intervals $T_{\rm m}^*$ and $T_{\rm M}^*$ of cycle 12\,--\,23 
  determined in Paper~I,
 we have found 
the following:
 
\begin{enumerate}
\item   $A_{\rm N} (T_{\rm m}^*)$  varies strongly with a period of  
 about 44 years (double Hale cycle),
 whereas there exits a weak 80\,--\,90 year Gleissberg cycle
  in $A_{
\rm S} (T_{\rm m}^*)$. 

\item   $A_{
\rm S} (T_{\rm M}^*)$ is also having a 44-year periodicity, 
  whereas $A_{\rm N} (T_{\rm M}^*)$  
 has  a somewhat different periodicity of 
about 33\,--\,44 years.

\item  There exist  statistically significant north-south asymmetries 
in the aforementioned sums of the areas of spot groups, $i.e.$, there exist
 statistically significant   differences  
${\delta A} (T_{\rm m}^*) = A_{\rm N} (T_{\rm m}^*) - A_{
\rm S} (T_{\rm m}^*)$ and ${\delta A} (T_{\rm M}^*) = A_{\rm N} (T_{\rm M}^*) - A_{
\rm S} (T_{\rm M}^*)$.

\item   ${\delta A} (T_{\rm m}^*)$ and ${\delta A} (T_{\rm M}^*) $
  vary approximately in opposite phase. 
 Both these vary strongly with a periodicity of about   44 years. 

\item  The patterns of the 44-year  cyclic  variations in  
  ${\delta A} (T_{\rm m}^*)$
 and ${\delta A} (T_{\rm M}^*)$  
strongly  indicate that the upcoming  cycle~24 will 
be weaker than cycle~23 and it will be the beginning minimum 
epoch of the next 44-year cycle (it may be also  the end of the 
current Gleissberg cycle). The next such a minimum may be 
represented by cycle~28.  

\item  As in the case of   $A_{\rm N} (T_{\rm m}^*)$ and $A_{
\rm S} (T_{\rm M}^*)$,
 the  north-south difference, ${\delta A} (T_{\rm m}^*)$   
 of a cycle,  
 correlates 
well with $R_{\rm M}$ of its immediate following cycle. 
By using this relationship it is possible to predict 
the amplitude of a cycle by about 13-year 
advance,  with an improved accuracy.
Using this relationship 
 we have obtained $103 \pm 10$ for 
the amplitude of the upcoming solar cycle~24.

\item  The correlation between the sum 
 ${\delta} A (T_{\rm m}^*)  + A_{\rm S}(T_{\rm M}^*)$ 
 of a cycle  and $R_{\rm M}$ of its immediate  following cycle is very high.    
By using this relationship it is possible to predict
 the amplitude of a cycle 
by about 9 years in advance with a high accuracy.
Using this relationship 
 we have obtained $87 \pm 7$ for $R_{\rm M}$ of cycle~24, 
which is about 28\% less than the $R_{\rm M}$ of cycle~23. 

\item In the cycle-to-cycle variations 
of $A_{\rm N} (T_{\rm m}^*)$ and $A_{
\rm S} (T_{\rm M}^*)$ there is  an indication  of  
 the upcoming cycle pair~24,25 
will not violate the G-O rule, $i.e.$ it seems cycle~25 will be stronger than 
cycle~24.  

\item There exists a considerable difference between the variations in the mean 
meridional motions of the spot groups during the odd numbered and the even numbered cycle.  
These variations suggest that there exist 
 about 9\,--\,10 year and 15\,--\,20 year periodicities 
in the mean meridional motions of the spot groups
in $0^\circ -10^\circ$  
  latitude intervals of the  
southern and the northern hemispheres, respectively.

\item  The spacial (latitudinal)
 patterns of the aforementioned  variations in the mean  
meridional motions of the sunspot groups suggest that the solar  meridional 
flow may transport  
 magnetic flux across the solar equator and 
potentially responsible for  all the relationships  
found here. 
Consequently, our results are not only highly statistically significant 
and useful 
for an accurate prediction of the amplitude 
of the upcoming cycle~24, but also have many implications for understanding 
the underlying physical process responsible for the solar cycle. 
\end{enumerate}

\begin{acknowledgements}
I thank the  anonymous referee for useful comments and suggestions.
 I also thank   
 Professor Roger K. Ulrich 
   for fruitful discussion,  and  Dr. R. W. Komm,
 Dr. D. H. Hathaway and Dr. K. M. Hiremath 
 for useful comments.  I  
 acknowledge the funding  by NSF grant AM-0236682.
\end{acknowledgements}

\end{article}
\end{document}